\documentclass[useAMS,usenatbib]{mn2e}
\usepackage{graphicx}
\usepackage{longtable}

\title[Ejection velocity fields of asteroid
families]
{Constraints on the original ejection velocity fields of asteroid families}
\author[V. Carruba, D. Nesvorn\'{y}]{V. Carruba$^{1,2}$\thanks{E-mail: vcarruba@feg.unesp.br}, D. Nesvorn\'{y}$^{1}$\\
$^{1}$Department of Space Studies, Southwest Research Institute, Boulder, 
CO, 80302, USA.\\
$^{2}$UNESP, Univ. Estadual Paulista, Grupo de din\^{a}mica Orbital e
  Planetologia, Guaratinguet\'{a}, SP, 12516-410, Brazil \\
}

\begin{document}

\date{Accepted January 5 2016.  Received December 22 2015; in original form November 27 2015.}

\pagerange{\pageref{firstpage}--\pageref{lastpage}} \pubyear{2015}

\maketitle

\label{firstpage}

\begin{abstract}
Asteroid families form as a result of large-scale collisions among
main belt asteroids. The orbital distribution of fragments after a 
family-forming impact could inform us about their ejection velocities.  
Unfortunately, however, orbits dynamically evolve by a number of effects, 
including the Yarkovsky drift, chaotic diffusion, and gravitational 
encounters with massive asteroids, such that it is difficult to infer 
the ejection velocities eons after each family's formation. 
Here we analyze the {\it inclination} distribution of asteroid 
families, because proper inclination can remain constant over long 
time intervals, and could help us to understand the distribution of 
the component of the ejection velocity that is perpendicular to the 
orbital plane ($v_W$).  From modeling the initial breakup, we find 
that the distribution of $v_W$ of the fragments, which manage to escape 
the parent body's gravity, should be more peaked than a Gaussian 
distribution (i.e., be leptokurtic) even if the initial distribution was 
Gaussian. We surveyed known asteroid families for signs of a peaked 
distribution of $v_W$ using a statistical measure of the distribution 
peakedness or flatness known as kurtosis. We identified eight families 
whose $v_W$ distribution is significantly leptokurtic. These cases (e.g. 
the Koronis family) are located in dynamically quiet regions of the 
main belt, where, presumably, the initial distribution of $v_W$ was not 
modified by subsequent orbital evolution. We suggest that, in these cases, 
the inclination distribution can be used to obtain interesting information 
about the original ejection velocity field.
\end{abstract}

\begin{keywords}
Minor planets, asteroids: general -- celestial mechanics.  
\end{keywords}
%

\section{Introduction}
\label{sec: intro}
Asteroid families form as a result of collisions between asteroids.
These events can either lead to a formation of a large crater 
on the parent body, from which fragments are ejected, or catastrophically 
disrupt it.  More than 120 families are currently known in the main
belt (Nesvorn\'{y} et al. 2015) and the number of their members ranges
from several thousands to just a few dozens, for the smaller and compact 
families.  A lot of progress has been made in the last decades in developing
sophisticated impact hydrocodes able to reproduce the main properties
of families, mainly accounting for their size distribution, and, in a few
cases (the Karin and Veritas clusters) the ejection velocities of their 
members (Michel et al. 2015).  However, while the sizes of asteroids can 
be either measured directly through radar 
observations or occultations of stars, or inferred if the geometric
albedo of the asteroid is known, correctly assessing ejection velocities
is a more demanding task.  The orbital element distribution
of family members can, at least in principle, be converted into 
ejection velocities from Gauss' equations (Zappal\`{a} et al. 1996), 
provided that both the true anomaly and the argument of perihelion of 
the family parent body are known (or assumed).  

Orbital elements of family members, however, are not fixed in time,
but can be changed by gravitational and non-gravitational 
effects, such as resonant dynamics (Morbidelli and Nesvorn\'{y} 1999), 
close encounters with massive asteroids (Carruba et al. 2003), and 
Yarkovsky (Bottke et al. 2001) effects, etc.   Separating which part of 
the current distribution in proper elements may be caused by the 
initial velocity field and which is the consequence of later evolution 
is a quite complex problem.   Interested readers are referred to 
Vokrouhlick\'{y} et al. (2006a,b,c) for a discussion of Monte Carlo 
methods applied to the distribution of asteroid families proper semi-major 
axis.  Yet, insights into the distribution of the ejection velocities are 
valuable for better understanding of the physics of large-scale collisions 
(Nesvorn\'{y} et al. 2006, Michel et al. 2015).  They may help to calibrate 
impact hydrocodes, and improve models of the internal structure of asteroids.
    
Here we analyze the inclination distribution of asteroid families.  The 
proper inclination is the the proper element least affected by 
dynamical evolution, and it could still bear signs of the original 
ejection velocity field.  We find that a family formed in 
an event in which the ejection velocities were not much larger than 
the escape velocity from the parent body should be characterized by a 
peaked (leptokurtic) initial distribution (relative to a Gaussian), 
while families formed in hyper-velocity impacts, such 
as the case of the Eos family (Vokrouhlick\'{y} et al. 2006a), should
have either a normal or less peaked (platykurtic) distribution.  
The subsequent dynamical evolution should then act to bring this 
initial distribution to appear more Gaussian (or mesokurtic). 
The relative importance of the subsequent evolution depends on which 
specific proper element is considered, and on how active the local 
dynamics is.  Using the proper inclination we attempt to identify cases 
where the local dynamics either did not have time or was not effective in
erasing the initial, presumably leptokurtic, distributions.  These cases 
can be used to better understand the conditions immediately after a 
parent body disruption.

This paper is divided as follows. In Sect.~\ref{sec: gauss_eq}, we model
the distribution of ejection velocities after a family-forming event.
We explain how a peakedness of an expected distribution can be measured by 
the Pearson kurtosis. Sect.~\ref{sec: kurt_dyn} shows how dynamics can 
modify the initial distribution by creating a new distribution that is 
more Gaussian in shape.  In Sect.~\ref{sec: kurt_families}, we survey the 
known asteroid families, to understand in which cases the traces of the 
initially leptokurtic distribution can be found. Sect.~\ref{sec: conc} 
presents our conclusions.
\section{Model for the initial $v_W$ distribution}
\label{sec: gauss_eq}

Proper orbital elements can be related to the components of 
the velocity in infinity, ${\vec{v}}_{inf}$, along the direction of 
the orbital motion ($v_t$), in the radial direction ($v_r$), and 
perpendicular to the orbital plane ($v_W$) through the Gauss equations 
(Murray and Dermott 1999):

\begin{equation}
\frac{\delta a}{a} = \frac{2}{na(1-e^2)^{1/2}}[(1+e\cos{(f)} \delta v_t +
(e \sin{(f)}) \delta v_{r}],
\label{eq: gauss_1}
\end{equation}

\begin{equation}
\delta e =\frac{(1-e^2)^{1/2}}{na}\left[\frac{e+cos(f)+e cos^2 (f)}{1+e cos (f)}
\delta v_t+sin(f) \delta v_r\right],
\label{eq: gauss_2}
\end{equation}

\begin{equation}
\delta i = \frac{(1-e^2)^{1/2}}{na} \frac{cos(\omega+f)}{1+e cos(f)} \delta v_W. 
\label{eq: gauss_3}
\end{equation}

\noindent 
where $\delta a = a-a_{ref}, \delta e = e-e_{ref}, \delta i= i-i_{ref}$, 
and $a_{ref}, e_{ref}, i_{ref}$ define a reference orbit (usually the center 
of an asteroid family, defined in our work as the center of mass
in a 3D proper element orbital space~\footnote{For families formed in 
cratering events, the center of mass of the asteroid family usually 
coincides with the orbital position of the parent body. One notable
exception to this rule is the Hygiea family, for reasons discussed
in Carruba et al. (2014), such as possible dynamical evolution caused
by close encounters with massive bodies with 10 Hygiea itself after
the family formation.},  
and $f$ and $\omega$ are the (generally unknown) true 
anomaly and perihelion argument of the disrupted body at the time of impact.  
From the initial distribution $(a,e,i)$, it should therefore be possible to 
estimate the three velocity components, assuming one knows the values of 
$f$ and $\omega$.  This can be accomplished by inverting 
Eqs.~\ref{eq: gauss_1},~\ref{eq: gauss_2},~\ref{eq: gauss_3}.  With the 
exception of extremely young asteroid families (e.g., 
the Karin cluster, Nesvorn\'{y} et al. 2002, 2006), this approach is not 
viable in most cases.  Apart from the obvious limitation that 
we do not generally know the values of $f$ and $\omega$, a more fundamental
difficulty is that several gravitational (e.g., mean-motion and secular 
resonances, close encounters with massive asteroids) and non-gravitational 
(e.g., Yarkovsky and YORP) effects act to change the proper elements on 
long timescales. The Gauss equations thus cannot be directly applied in 
most cases to obtain information about the original ejection velocities.

In this work we propose a new method to circumvent this difficulty.  Of the 
three proper elements, the proper inclination is the one that is the least 
affected by dynamics.   For example, unlike the proper semi-major axis, the 
proper inclination is not directly effected by the diurnal version of the 
Yarkovsky effect (it is affected by the seasonal version, but this
is usually a much weaker effect, whose strength is of the order of 10\%
of that of the diurnal version, for typical values of asteroid
spin obliquities and rotation periods (Vokrouhlick\'{y} and Farinella 1999)).  
Also, unlike the proper eccentricity, which is affected by 
chaotic diffusion in mean-motion resonances, the proper inclination is more 
stable.  In addition, the inclination is related  to a single component of 
the ejection velocities, $v_W$, via Eq.~\ref{eq: gauss_3}.  This equation can 
be, at least in principle, inverted to provide information about 
$\delta v_W$.  

\begin{figure*}

  \centering
  \begin{minipage}[c]{0.5\textwidth}
    \centering \includegraphics[width=2.5in]{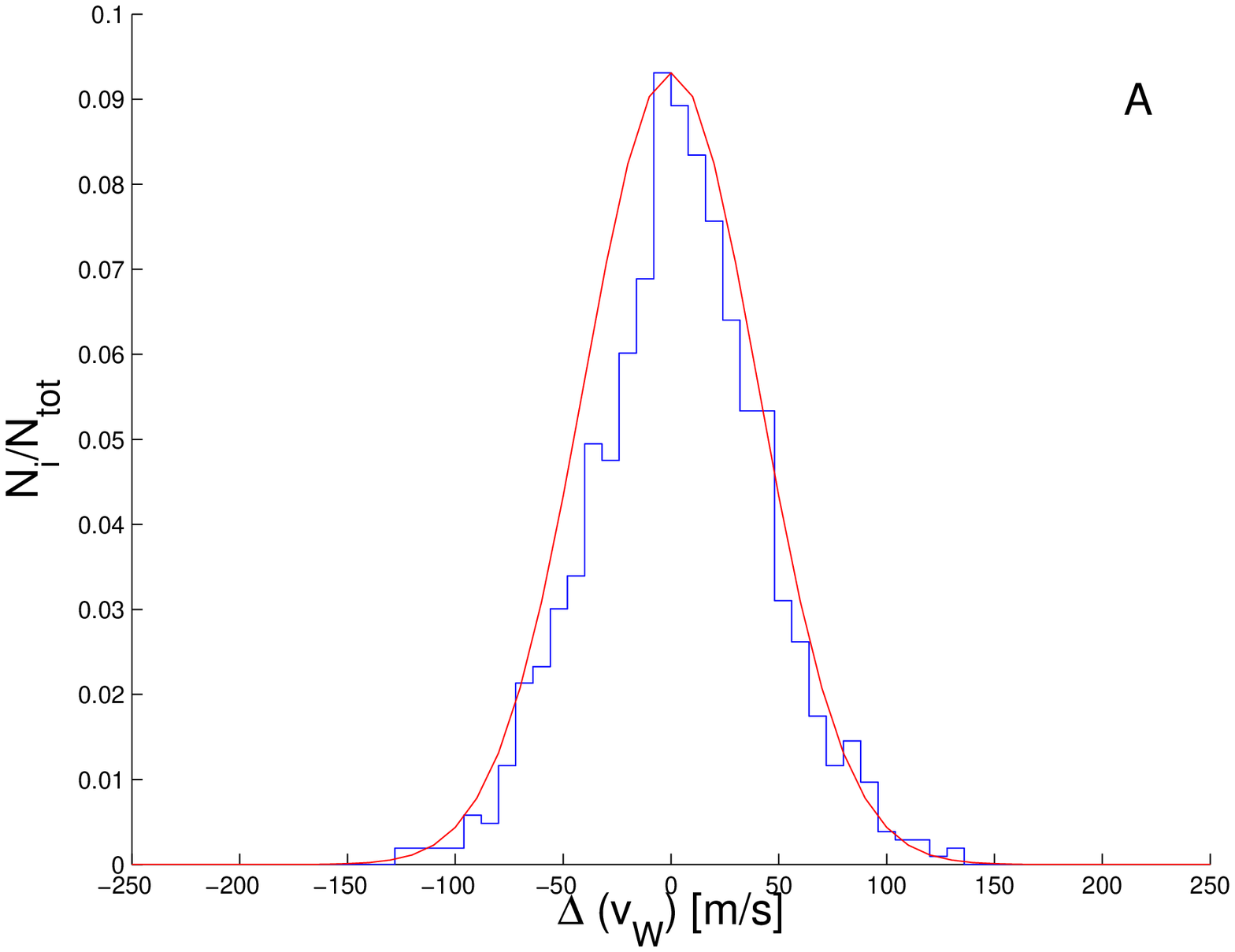}
  \end{minipage}%
  \begin{minipage}[c]{0.5\textwidth}
    \centering \includegraphics[width=2.5in]{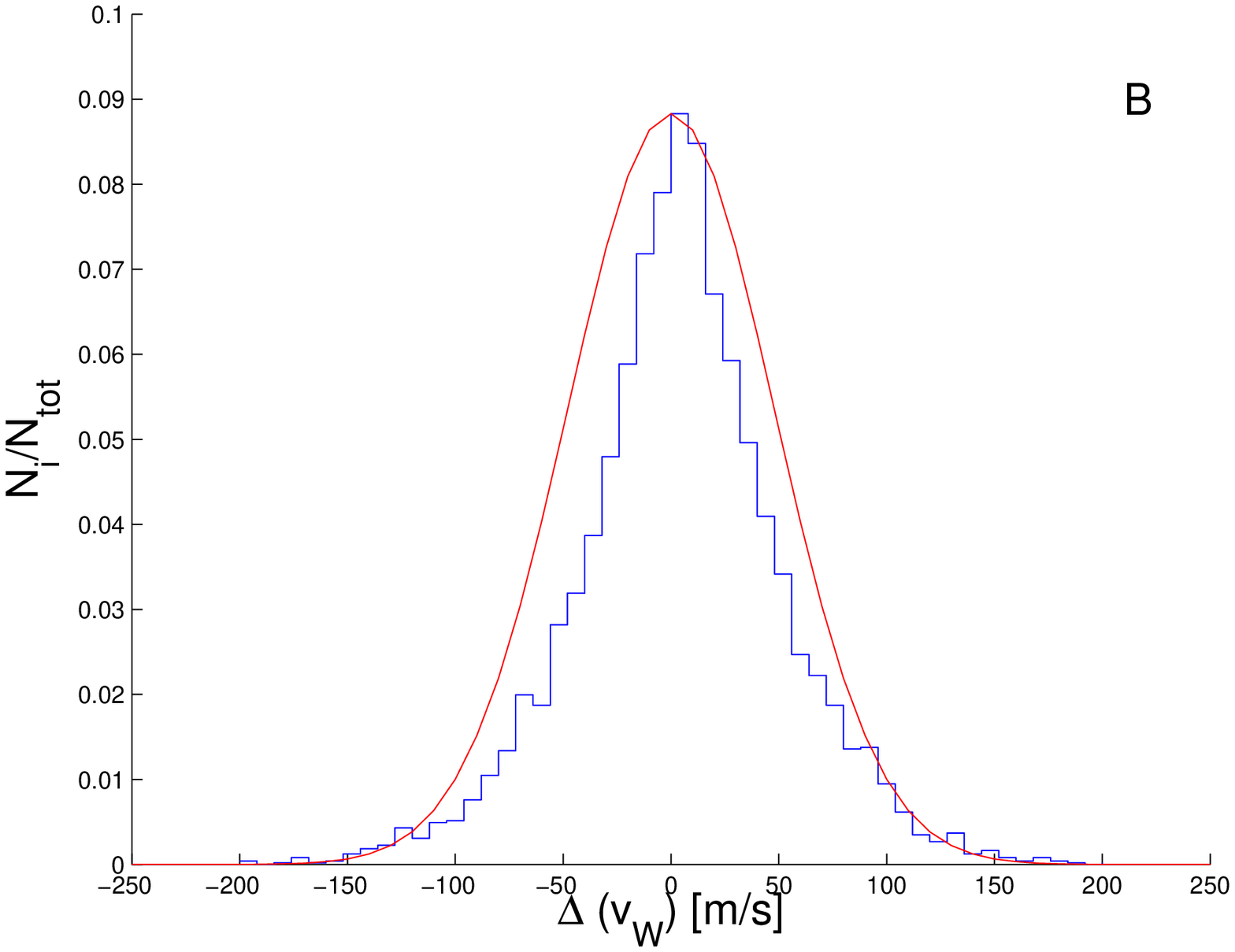}
  \end{minipage}

\caption{The $v_W$ distribution for a simulated family. Panel A shows the 
distribution for objects with diameters $2.5 < D < 3.5$ km. Panel B shows 
the results for $2.0 < D < 8.0$~km. The size distribution was set to 
be $N\,{\rm d}N = C D^{-\alpha}\,{\rm d}N$ with $\alpha=3.6$.  We also 
set $V_{EJ}/V_{esc} = 0.5$.  A Gaussian distribution with the 
same standard deviation of the $v_W$ distribution is shown in both panels 
for reference.}
\label{fig: beta_fam_real}
\end{figure*}

What kind of a probability distribution function ($pdf$) is to be expected
for the original values of $v_{w}$?  Velocities at infinity $V_{inf}$
are obtained from the ejection velocities $V_{ej}$ through the relationship:

\begin{equation}
V_{inf}=\sqrt{V_{ej}^2 -V_{esc}^2},
\label{eq: vel_inf}
\end{equation}

\noindent
where $V_{esc}$ is the escape velocity from the parent body.  Following
Vokrouhlick\'{y} et al. (2006a,b,c), we assume that the ejection velocity
field follows a Gaussian distribution of zero mean and standard deviation
given by:

\begin{equation}
{\sigma}_{V_{ej}}=V_{EJ}\cdot \frac{5km}{D},
\label{eq: std_veJ}
\end{equation}

\noindent 
where $D$ is the body diameter in km, and $V_{EJ}$ is a parameter 
describing the width of the velocity field.  Only objects with 
$V_{ej} > V_{esc}$ succeed in escaping from the parent body.  As 
a result, the initial distribution of $V_{inf}$ should generally be 
more peaked than a Gaussian one.  Fig.~\ref{fig: beta_fam_real}, panel A, 
displays the frequency distribution function $fdf$ for 1031 objects with 
$2.5 < D < 3.5$ km computed for a parent body with an escape velocity of 
130 m/s and $V_{EJ} = 0.5 V_{esc}$.  We assumed that 
$f=30^{\circ}$ and $(\omega +f) =50^{\circ}$.  Since $f$ and $(\omega +f)$ appear 
only as multiplying factors in the expression for $v_W$, different choices 
of these parameters do not affect the shape of the distribution.   One can 
notice how the $v_W$ distribution is indeed more peaked than that 
of the Gaussian distribution with the same standard deviation (shown in 
red in Fig.~\ref{fig: beta_fam_real}).

A useful parameter to understand if a given distribution is (or is not) 
normally distributed is the kurtosis (Carruba et al. 2013a).  Pearson kurtosis 
(Pearson 1929), defined as ${\gamma}_{2}= \frac{{\mu}_4}{{\sigma}^{4}}-3$, 
where ${\mu}_4$ is the fourth moment about the mean and $\sigma$ is the 
standard deviation, gives a measure of the ``peakedness'' of the distribution.
The Gaussian distribution has ${\gamma}_2 = 0$ and is the most commonly 
known example of a mesokurtic distribution.  Larger values of ${\gamma}_2$ 
are associated with leptokurtic distributions, which have longer tails and 
more acute central peaks. The opposite case, with ${\gamma}_2 < 0$, is 
known as platykurtic.

\begin{figure}
\centering
\centering \includegraphics [width=0.45\textwidth]{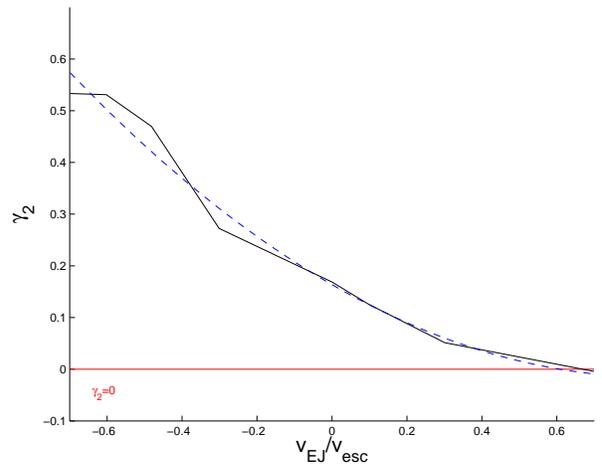}
\caption{Dependence of the kurtosis ${\gamma}_2$ of the $v_W$ distribution 
on $V_{EJ}/V_{esc}$ (solid blue line).  The scaling of the tick marks 
on the x-axis are proportional to the $\log_{10}{V_{EJ}/V_{esc}}$.
The blue dashed line is the second order polynomial that best-fitted 
the data.  The red line is ${\gamma}_2 = 0$, corresponding to a mesokurtic 
distribution.}
\label{fig: beta_vej}
\end{figure}

We generated values of $v_W$ for various values of $V_{EJ}$ for ten thousand
5-km bodies originating from a parent body with $V_{esc}=130$~m~s$^{-1}$ 
(results can be re-scaled for any other body size using Eq.~\ref{eq: std_veJ}, 
and a different value of $D$).  Fig.~\ref{fig: beta_vej} shows how the 
kurtosis value of the $v_W$ distribution changes as a function of 
$V_{EJ}/V_{esc}$.  As expected, for smaller values of $V_{EJ}/V_{esc}$ the 
$v_W$ distribution is more peaked, and the kurtosis values are larger.  
For $V_{EJ}/V_{esc}>1$, the kurtosis value approaches 0, because the 
influence of parent body's gravity diminishes.  Most families for 
which an estimate of $V_{EJ}$ is available (see Nesvorn\'{y} et al. (2015), 
Sect.~5 and references within) have 
estimated values of $V_{EJ}/V_{esc}$ in the range from 0.4 to 1.2, and 
should therefore have been significantly leptokurtic when they formed.

What typical values of ${\gamma}_2$ one would expect for a real asteroid
family, where different sizes of fragments are considered together?  As the 
large objects typically have lower values of $v_W$, this implies that they 
would contribute to the peak of the distribution, and when considered with 
smaller fragments, the whole distribution should correspond to a larger 
value of ${\gamma}_2$.  To illustrate this effect, we simulated a family 
using a size distribution $N\,{\rm d}N = C D^{-\alpha}\,{\rm d}N$ with 
$\alpha=3.6$, and $V_{EJ}/V_{esc} = 0.5$.  The computed value of ${\gamma}_2$ 
for bodies in the size range from 2 to 8 km was found to be 0.96 (see 
Fig.~\ref{fig: beta_fam_real}, panels B), while the one for a restricted 
range $2.5 < D < 3.5$ km was 0.21 (Fig.~\ref{fig: beta_fam_real}, panel A).  
This shows that one must be careful when interpreting the real families, 
where the escape velocity and size distribution effects can combine 
together to produce larger values of ${\gamma}_2$.  To isolate the 
escape velocity effect, it is best to use a restricted range of sizes.    

Finally, we compare our simple model for the $v_W$ distribution with 
the results of impact simulations.  We have taken these results from 
Nesvorn\'{y} et al. (2006), where a Smooth Particle Hydrodynamic (SPH) 
code was used to model the formation of the Karin family.  In this case, 
the parent body was assumed to have $D = 33$~km,which gives 
$V_{ESC} \simeq 35$~m s$^{-1}$.  Different impact conditions (impact speed, 
impact angle, projectile size, etc.) were studied in this work.  The 
dynamical evolution of fragments and their re-accretion following the 
initial impact was followed by the PKDGRAV code.

Using these simulations, we extracted the values of $v_W$ for the escaping 
fragments and estimated the value of $V_{EJ}$ from Eq.~\ref{eq: vel_inf}.  In 
a specific case that produced the best fit to the observed size distribution 
of the Karin family, we obtained $V_{EJ}/V_{esc} =0.2$ for fragments with 
$1 < D < 5$~km. Finally, we found that the $v_W$ distribution has 
${\gamma}_2=0.5$, and is therefore significantly peaked.
This is in very good agreement with results from our simple method to 
generate fragments (see above), which for the same size range and 
$V_{EJ}/V_{esc}$ gives ${\gamma}_2=0.53$.  This suggests, again, that 
the initial $v_W$ distribution of asteroid families should 
be leptokurtic.
\section{Effects of long-term dynamics}
\label{sec: kurt_dyn}
Now that we have some expectations for the ${\gamma}_2$ values just after 
a family formation, we turn our attention to ${\gamma}_2$ for real 
asteroid families, where the inferred $v_W$ values may have been affected 
by the long-term dynamics.  We use the Koronis family to illustrate this 
effect. The Koronis family is located at low eccentricities and low 
inclinations, in a region that is relatively dynamically quiet (at least 
in what concerns the proper inclination; Bottke et al. 2001).

\begin{figure}
\centering
\centering \includegraphics [width=0.45\textwidth]{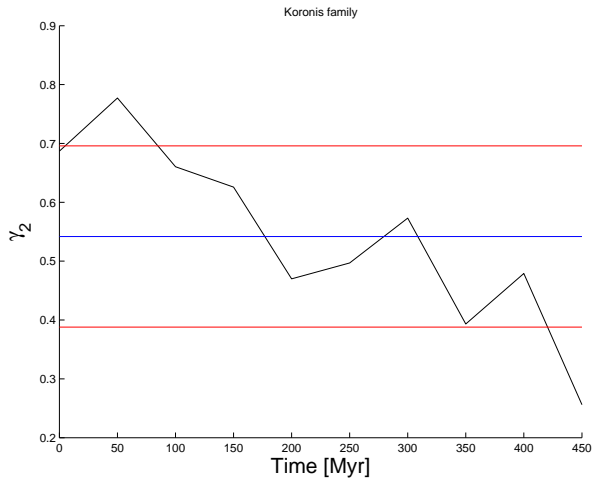}
\caption{The time evolution of the kurtosis for a simulated Koronis family
(Bottke et al. 2001). As described in the main text, we computed ${\gamma}_2$
of the $v_W$ component and $D=2$ km bodies (black line). The blue line 
shows the mean value over the 450-Myr integration span, and the red lines 
are one standard deviation from the mean.}
\label{fig: kurtosis_vW}
\end{figure}

Bottke et al. (2001) simulated the long-term dynamical evolution of the Koronis 
family. Their simulation included the gravitational effects of four outer 
planets and the Yarkovsky effect. Different sizes of simulated family members  
were considered. We opted to illustrate here the case with $D=2$ km, for which
Bottke et al. (2001) numerically integrated the orbits of 210 bodies over 
450 Myr.  The initial distribution of family members in Bottke et al. (2001) 
was obtained using the same method as described in Sect.~\ref{sec: gauss_eq}.  
We computed the values of $v_{w}$ by inverting Eq.~\ref{eq: gauss_3}.  Since 
Pearson's kurtosis is dependent on the presence of outliers, we eliminated 
from our distributions objects with values more than $4 \sigma$ away from 
the mean. This allowed us to avoid possible distortions in the computed 
${\gamma}_2$ values caused by a few distant objects. 

Fig.~\ref{fig: kurtosis_vW} shows that the initially leptokurtic distribution 
tends to become more mesokurtic with time. This can be understood as follows.  
In statistics, the central limit theorem states that the averages of random 
variables drawn from uncorrelated distributions are normally distributed.  If 
the dynamical effects produce changes of $i_P$ compatible with the central 
limit theorem, one would expect that, with time, the distributions of 
$v_{w}$ should indeed become more and more mesokurtic, as the 
contribution from dynamics increases.

To illustrate things in the case of the (real) Koronis family, we computed 
the value of ${\gamma}_2$ for all its 5118 members taken from Nesvorn\'{y} 
et al. (2015), and for members with $4.5 < D < 5.5$~km.  We obtained  
${\gamma}_2 = 0.993$ and $0.712$, respectively.  Since both these values 
are considerably leptokurtic, we believe that the dynamical evolution of 
the Koronis family did not had enough time to alter the initial distribution of 
inclinations. Therefore, in the specific case of the Koronis family, we may 
still be observing traces of the primordial distribution of $v_W$.
\section{Kurtosis of the real asteroid families}
\label{sec: kurt_families}
We selected the families identified in Nesvorn\'{y} et al. (2015) that: (i) 
have at least 100 members with $2 < D < 4$~km, and (ii) have a reasonably 
well defined age estimate\footnote{Families were determined in 
Nesvorn\'{y} et al. 2015 using the Hierarchical Clustering Method (HCM)
in the $(a,e,sin(i))$ proper element domain. This method may not identify 
peripheral regions of some large families, the so-called {\it halo} of 
Bro\v{z} and Morbidelli (2013), as belonging to the dynamical group.  
For families such as the Eos and Koronis
groups, results from Nesvorn\'{y} et al. (2015) should be considered as 
conservative estimates.}.  The first selection criterion is required to 
have a reasonable statistics for the ${\gamma}_2(v_W)$ computation.  If we 
assume that the error is proportional to the square root of the number of 
objects, a sample of 100 objects would give us an uncertainty in 
${\gamma}_2(v_W)$ of 10\%.  We use a restricted size range, $2 < D < 4$~km, 
to avoid the influence of the size-dependent velocity effects on the kurtosis.  
The age estimate is useful to have some input for a physical interpretation 
of our results.  Also, since ${\gamma}_2(v_W)$ is sensitive to the presence 
of outliers in the distribution, we use the Jarque-Bera statistical test 
(jbtest hereafter) to confirm that the $v_W$ distribution differs (or not) 
from a Gaussian one.  We choose to work with this specific test, instead of
others, because the jbtest is particularly sensitive to whether sample 
data have the skewness and kurtosis matching a normal distribution (Jarque 
\& Bera 1987). The test, implemented on MATLAB, provides a $p$ parameter 
stating the probability that a given $v_W$ distribution follows (or not) 
the Gaussian $pdf$.   The null hypothesis level is usually set at 5\%.  
Finally, we also checked that the $v_W$ distribution was not too 
asymmetrical, and verified that its skewness, the parameter that measures a
possible asymmetry, was in the range between -0.25 and 0.25. A skewness value 
outside this range would indicate an asymmetric ejection velocity field, 
or some dynamical effects that are beyond the scope of this study.

Of the 122 families listed in  Nesvorn\'{y} et al. (2015), 48 cases
satisfied these requirements.  At this stage of our analysis we do not 
eliminate interlopers and do not consider in detail the local dynamics.
Also, families with halos may have a further spread out distribution
in inclination than what accounted for by HCM, and therefore larger
values of ${\gamma}_2(v_W)$.  For these families, our results should be
considered as conservative.  Values of ${\gamma}_2(v_W)$ may be affected 
by these factors, but first we would like to see what families may be more 
interesting in terms of a simple first-order criteria.   
Table~\ref{table: families_kurt} reports 
the values of ${\gamma}_2(v_W)$ for the whole family (3rd column), for 
$2.0 < D < 4.0$~km ($D_3$) members (4rd column), and the $p_{jbtest}$ coefficient
of the jbtest for the $D_3$ population.  Note that the maximum value
of $p_{jbtest}$ is 50.0\%.   

Following the notation of Nesvorn\'{y} et al. (2015), the first two columns 
in Table~\ref{table: families_kurt} report the Family Identification Number 
(FIN) and the family name.  Finally, 
the sixth column displays the family age estimate, with its error.
The age estimate was obtained using Eq.~1 in Nesvorn\'{y} et al. (2015),
and values of the $C_0$ parameter, its error (also estimated
from Eq.~1 and the value of $\delta C_0$ from Nesvorn\'{y} et al. 
2015\footnote{The error on family ages should be considered as nominal,
since they do not account for the uncertainties on values of fundamental
parameters such as the asteroids bulk densities and thermal conductivities
(Masiero et al. 2012).}), and geometric albedos from Table 2 of that paper.  
Densities were taken from Carry (2012) for the namesake body of each 
family, when available.  If not, a standard value of 1.2~g cm$^{-3}$ for 
C-complex families and 2.5~g~cm$^{-3}$ for the S-complex families was 
used.  In a few cases, for which a better age estimate was available in 
the literature, we used these latter values. This includes the following 
cases (labeled by $\dagger$ in Table 1): Karin (Nesvorn\'{y} et al. 2002), 
Eos (Vokrouhlick\'{y} et al. 2006b), Hygiea (Carruba et al. 2013b), and 
Koronis, Themis, Meliboea, and Ursula (Carruba et al. 2015). Moreover, the 
families with values of skewness not in the range from -0.25 to 0.25 are 
identified by a ``(s)'' after family's name in Table 1.

To select families whose $v_W$ component may have not changed significantly 
with time, we adopted the following criteria: ${\gamma}_2(v_W)$ for the 
$D_3$ population has to be larger than 0.25, the estimated error on 
values of ${\gamma}_2(v_W)$, skewness in the range from -0.25 and 0.25, and 
$p_{jbtest}$ has to be lower than 5\%.  After this pre-selection was carried 
out, we were left with a sample of 9 candidates  This is much higher
than the 2 families one would expect to randomly fulfill this criteria, 
based on the number of families in our sample, and this suggests that a 
real effect might actually be observed in our data.
Ordered by a decreasing value of ${\gamma}_2(v_W(D_3))$, these 8 
candidates are the Gallia, Hoffmeister, 
Barcelona, Hansa, Massalia, Koronis, Rafita, Hygiea and Dora families.  We 
discarded the case of the Hoffmeister family, since this family is 
significantly affected by the ${\nu}_{1C}= g-g_{C}$ resonance with Ceres 
(Novakovi\'{c} et al. 2015).  

\begin{figure*}
  \centering
  \begin{minipage}[c]{0.5\textwidth}
    \centering \includegraphics[width=2.5in]{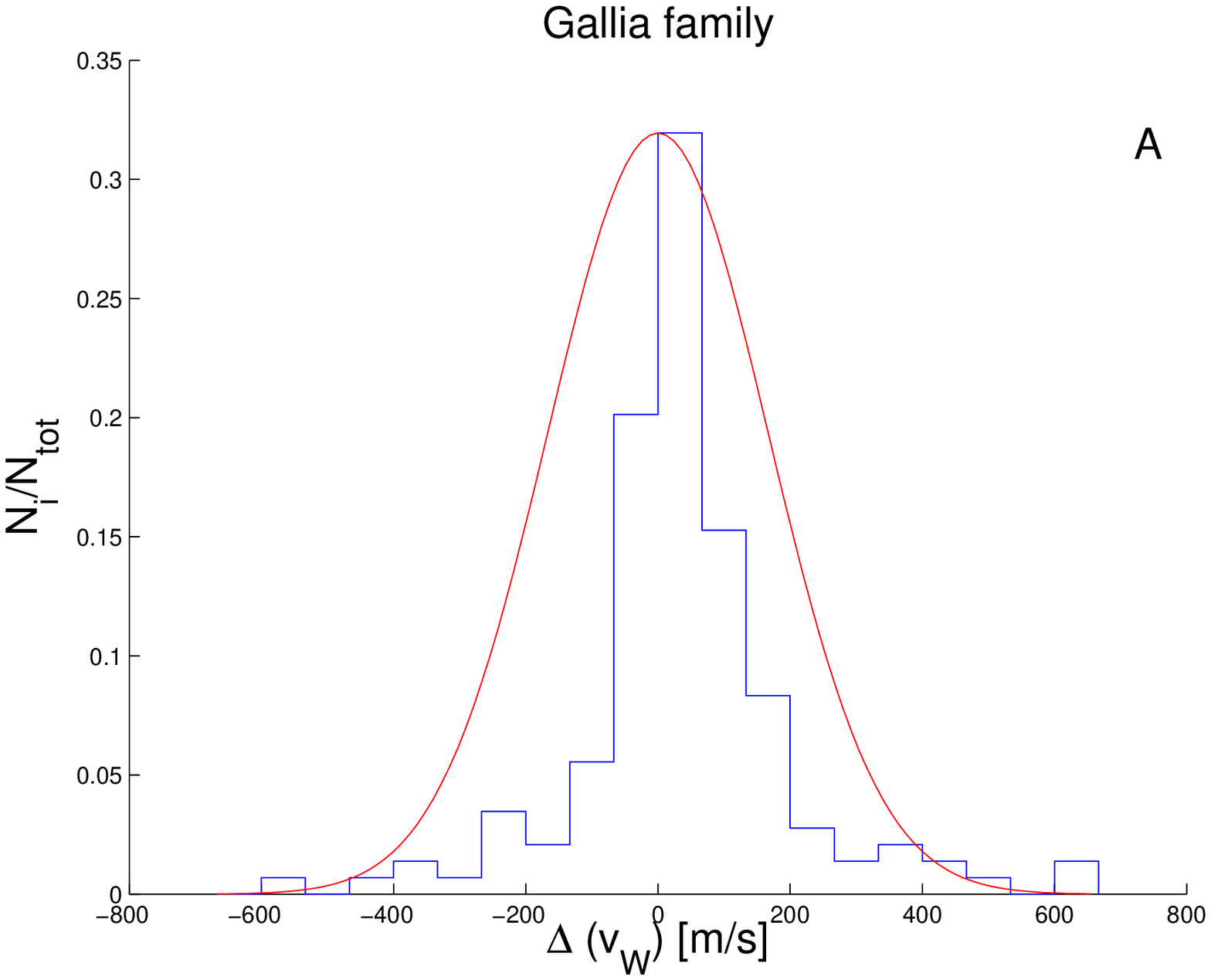}
  \end{minipage}%
  \begin{minipage}[c]{0.5\textwidth}
    \centering \includegraphics[width=2.5in]{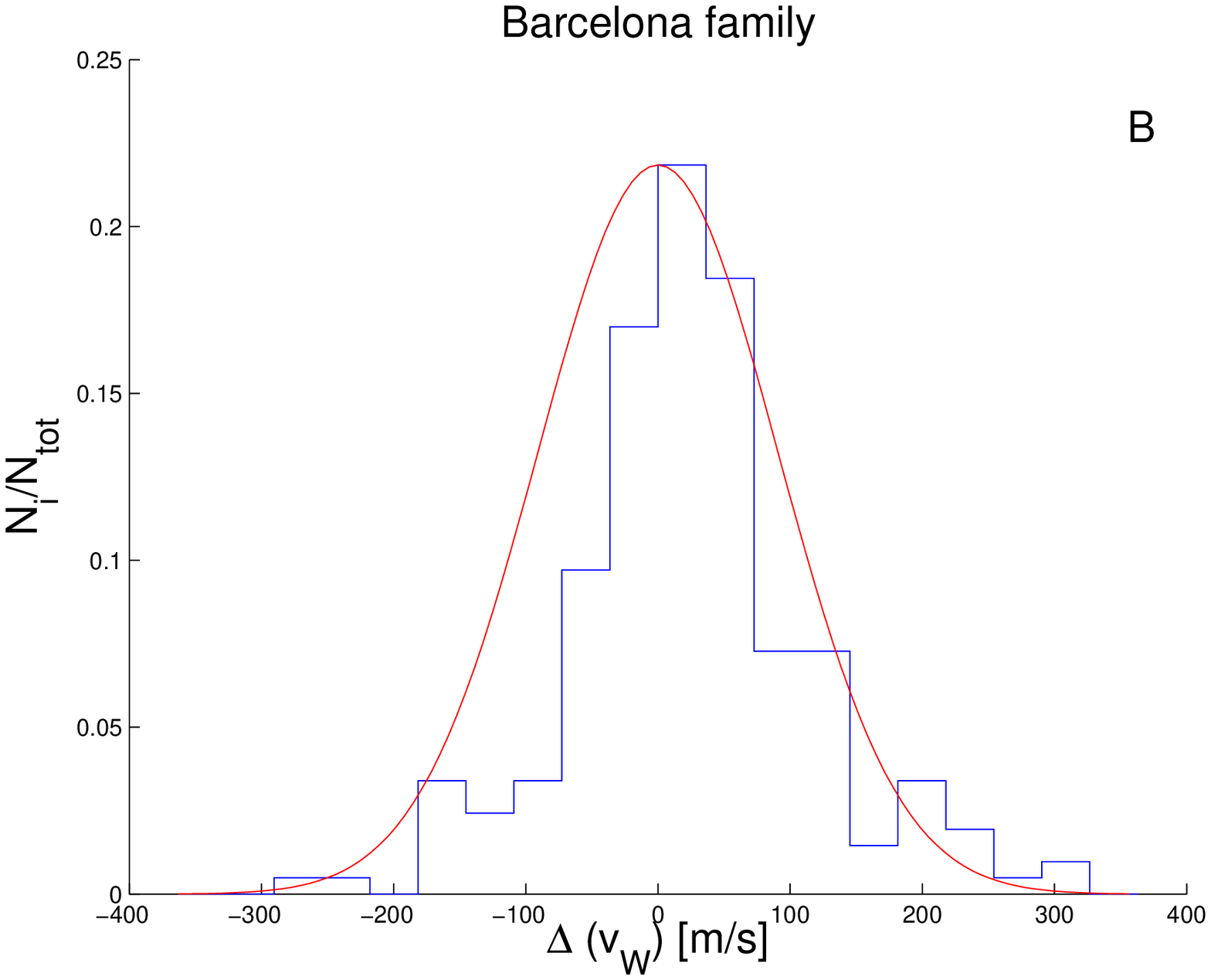}
  \end{minipage}
  \begin{minipage}[c]{0.5\textwidth}
    \centering \includegraphics[width=2.5in]{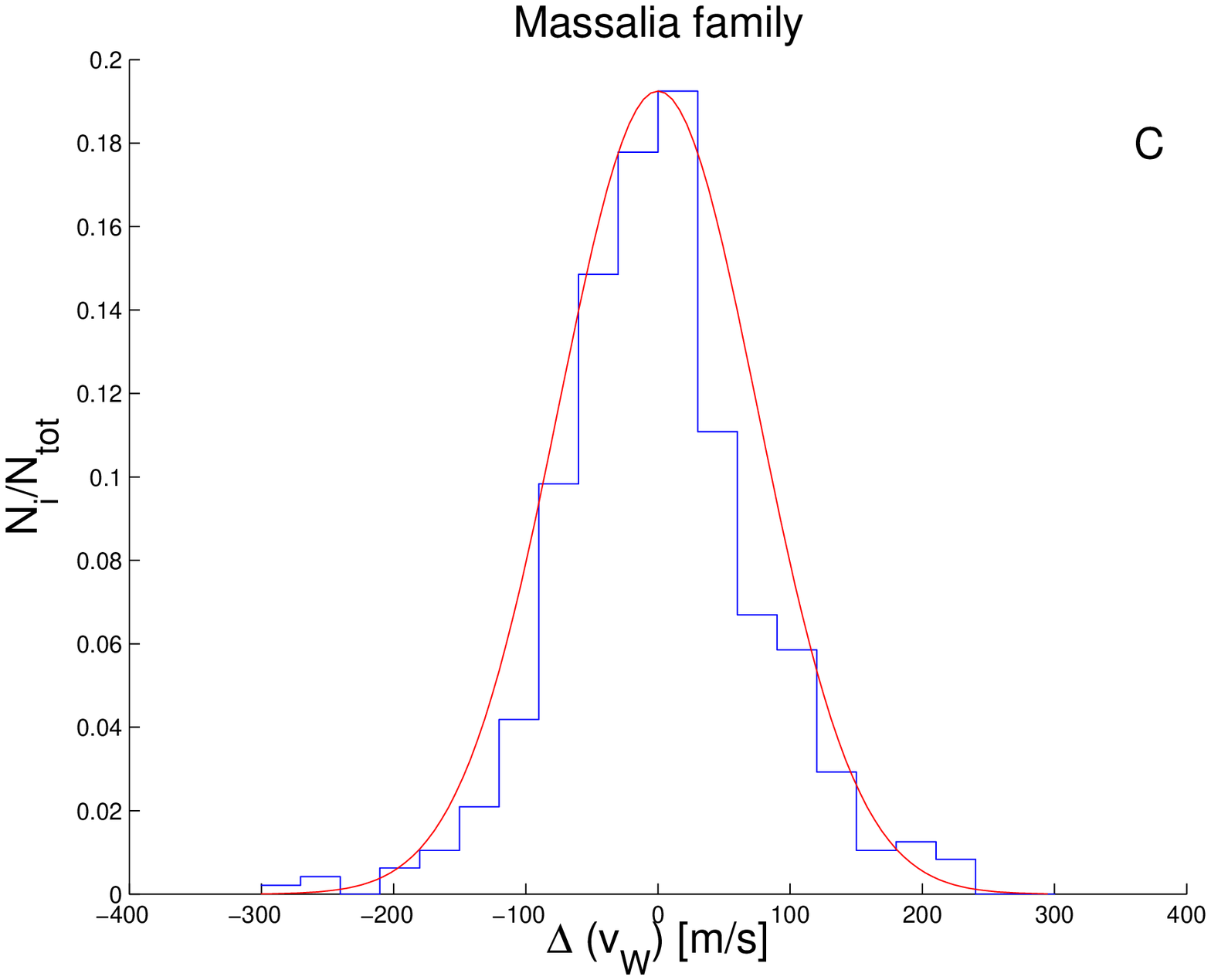}
  \end{minipage}%
  \begin{minipage}[c]{0.5\textwidth}
    \centering \includegraphics[width=2.5in]{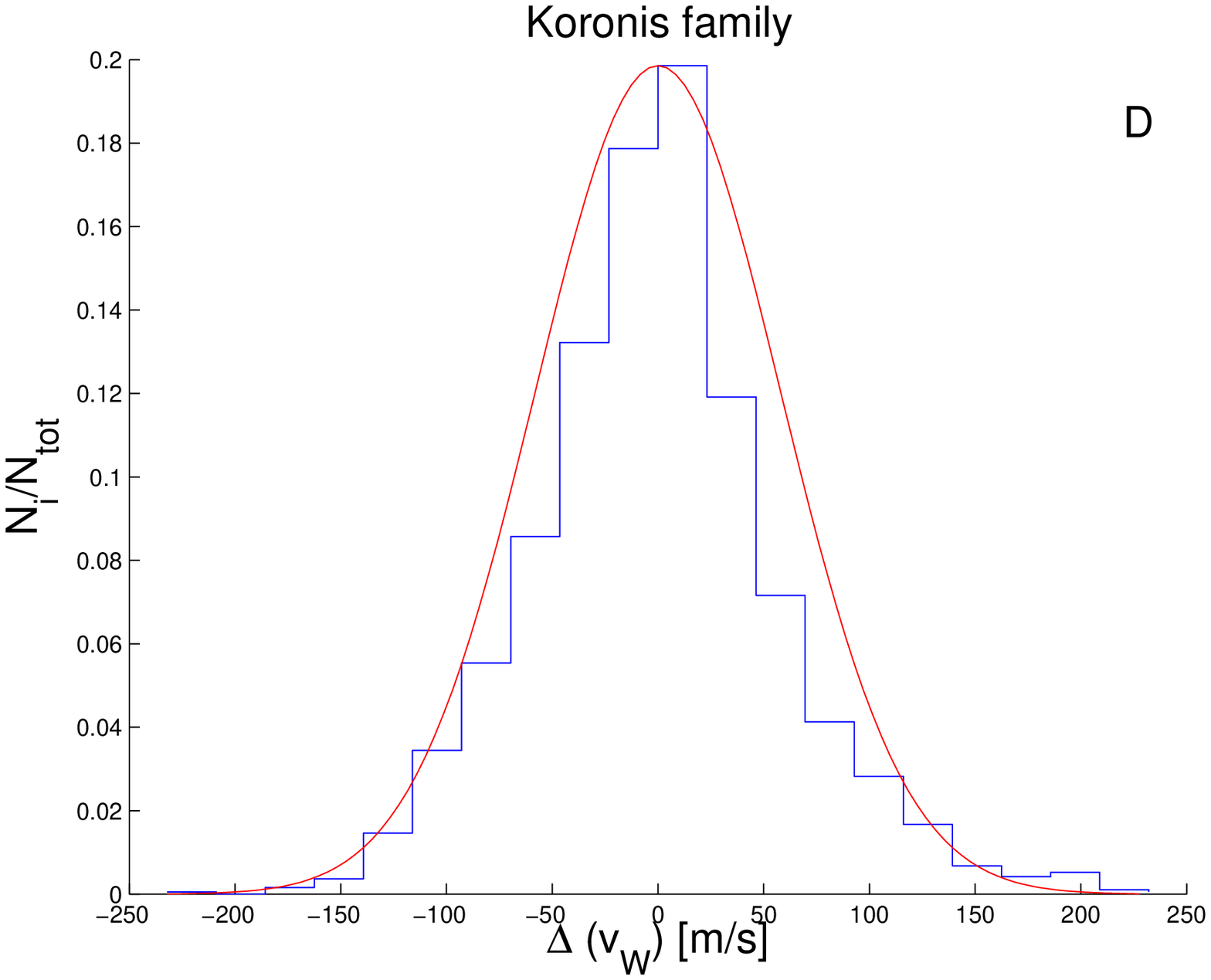}
  \end{minipage}
\caption{The $v_W$ distributions (blue line) of $D_3$ members of the Gallia 
(panel A), Barcelona (panel B), Massalia (panel C), and Koronis families 
(panel D). The red line displays the Gaussian distribution with the same 
standard deviation of the $v_W$ distribution and zero mean, normalized to 
the maximum value of the frequency distribution of each family.}
\label{fig: vW_famiies}
\end{figure*}

\begin{figure}
\centering
\centering \includegraphics [width=0.45\textwidth]{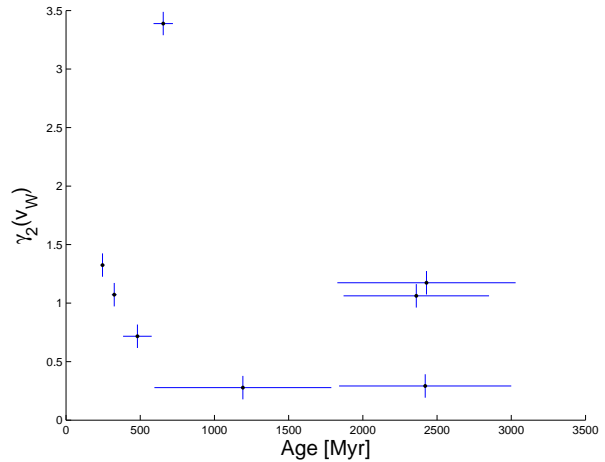}
\caption{${\gamma}_2(v_W(D_3))$ versus the estimated age of eight
families satisfying our selection criteria.  The horizontal blue 
lines display the estimated error on the ages, while the vertical
ones show the nominal errors on ${\gamma}_2(v_W(D_3))$, assumed equal
to 0.1.}
\label{fig: age_gamma}
\end{figure}

Fig.~\ref{fig: vW_famiies} shows the distributions (blue line) of the $v_W$ 
values for $D_3$ members of the Gallia, Barcelona, Massalia, and Koronis 
families, four of our selected families with leptokurtic distributions 
of $v_W$.  As can be observed from this figure, these families 
have $v_W$ more peaked than a Gaussian, and this is particularly 
evident for the case of the Gallia family.  Fig.~\ref{fig: age_gamma} 
displays values of ${\gamma}_2(v_W(D_3))$ versus the estimated ages of the 
eight families satisfying our selection criteria.  One can notice that there 
is not a clear correlation between family age and ${\gamma}_2(v_W(D_3))$.  
While there are 4 families younger than 700 Myr, 
three families (Hansa, Hygiea, and Rafita) are estimated to be older than 
2 Gyr.  What these eight families have in common is that they are located in 
orbital regions not very affected by dynamical mechanisms capable of 
changing proper $i$. Gallia, Barcelona, and Hansa are families in stable islands
at high inclinations, all in regions with not many mean-motion 
resonances (Carruba 2010).  As previously discussed,
Koronis is located in a region at low inclination relatively quiet in 
terms of dynamics, and the same can be said for Hygiea (Carruba
et al. 2013b), and, possibly, for the Rafita and Massalia families 
(the latter also a relatively young family, Vokrouhlick\'{y} et al. 2006b). 

Finally, among the families with large values of both skewness and kurtosis,
the cases of the Karin and Astrid families are of a particular
interest to us.  Karin is a very young family, already identified as such
in Nesvorn\'{y} et al. (2002, 2006), while Astrid is a family located
in a dynamically quiet region.  Nesvorn\'{y} et al. (2006) showed that the 
velocity field of Karin was probably asymmetrical, so the relatively
high value of the skewness that we found for the Karin cluster (-0.60) 
should not be surprising.  Astrid has the highest observed value of 
${\gamma}_2(v_W(D_3))$.  This family, however, also interacts with a 
secular resonance that produces a large dispersion of the inclination 
distribution.  If the resonant population of objects is removed, the value 
of ${\gamma}_2(v_W(D_3))$ drops to 0.3, with a skewness within the 
acceptable range.  The still large value of ${\gamma}_2(v_W(D_3))$ of 
this revised Astrid group makes it a very good candidate for a family with a 
partially pristine $v_W$ velocity field. 
\onecolumn
\begin{table*}
\begin{center}
\caption{Values of ${\gamma}_2(v_W)$ of the whole family 
(3rd column), the $2.0 < D < 4.0$~km ($D_3$) members (4rd column), 
the $p$ coefficient of the jbtest (5th column), and estimated family age
with its error (6th column) for the families in Nesvorn\'{y} et al. (2015)
that satisfy our selection criteria.  Families with values of skewness 
not in the range from -0.25 to 0.25 are identified by a ``(s)'' after the
family's name. Daggers identify the cases for which a more refined 
age estimate was available in the literature.}
\label{table: families_kurt}
\begin{tabular}{|c|c|c|c|c|c|}
\hline
FIN   & Family & ${\gamma}_2(v_W)$ & ${\gamma}_2(v_W)$ & $p_{jbtest}$ &   Age  \\
      & Name   &      All          & $D_3$             &   (\%)     &   [Myr]\\
\hline
      & Name   &      All          & $D_3$            &   (\%)     &   [Myr]\\
003 & 434 Hungaria (s)&0.42 & 0.21 & 0.5 & $135\pm13$    \\ 
401 &   4 Vesta     &-0.42 &-0.27 & 0.1 & $1440\pm720$  \\
402 &   8 Flora     &-0.59 &-0.60 & 0.1 & $4360\pm2180$ \\    
403 & 298 Baptistina&-0.03 &-0.16 & 0.1 & $110\pm10$    \\
404 &  20 Massalia  & 0.26 & 1.07 & 0.1 & $320\pm20$    \\
405 &  44 Nysa (s)  & 0.38 & 0.74 & 0.1 & $1660\pm80$   \\
406 & 163 Erigone   &-0.32 &-0.42 & 0.8 & $190\pm25$    \\
408 & 752 Sulamitis (s)&1.28& 1.56 & 0.1 & $320\pm30$    \\ 
413 &  84 Klio (s)  & 0.12 & 0.19 & 2.5 & $960\pm250$   \\
701 &  25 Phocaea   &-0.41 &-0.40 & 1.4 & $1460\pm1460$ \\
501 &   3 Juno (s)  & 1.05 & 1.05 & 0.1 & $740\pm150$   \\
502 &  15 Eunomia   &-0.42 &-0.44 & 0.8 & $3200\pm2240$ \\
504 & 128 Nemesis   & 0.19 & 0.20 & 3.5 & $440\pm20$    \\
505 & 145 Adeona    & 0.20 & 0.05 &19.2 & $620\pm190$   \\
506 & 170 Maria (s) &-0.17 &-0.38 & 0.1 & $1950\pm1950$ \\
507 & 363 Padua     &-0.28 &-0.18 & 4.8 & $410\pm80$    \\ 
510 & 569 Misa      & 0.14 & 0.02 &20.3 & $700\pm140$   \\
511 & 606 Brangane (s)&0.48&-0.48 & 3.2 & $60\pm5$      \\
512 & 668 Dora      & 0.16 & 0.28 & 1.0 & $1190\pm600$  \\
513 & 808 Merxia    & 0.24 & 0.34 & 9.9 & $340\pm30$    \\
514 & 847 Agnia (s) & 0.47 & 0.24 & 0.3 & $200\pm10$    \\
515 &1128 Astrid (s)& 3.69 & 4.83 & 0.1 & $140\pm10$    \\
516 &1272 Gefion    & 0.36 & 0.16 &11.9 & $980\pm290$   \\
517 &3815 Konig     & 0.94 & 0.48 &27.0 & $70\pm10$     \\
518 &1644 Rafita    & 0.63 & 0.72 & 0.3 & $480\pm100$   \\
519 &1726 Hoffmeister&2.21 & 1.82 & 0.1 & $270\pm10$    \\
533 &1668 Hanna     & 0.15 & 0.15 &19.3 & $240\pm20$    \\
535 &2732 Witt      &-0.22 &-0.21 &50.0 & $790\pm200$   \\
536 &2344 Xizang (s)& 2.47 & 1.87 & 0.1 & $220\pm20$    \\
539 & 369 Aeria (s) &-0.56 &-0.21 & 1.6 & $180\pm20$    \\
802 & 148 Gallia    & 2.02 & 3.39 & 0.1 & $650\pm60$    \\
803 & 480 Hansa     & 0.81 & 1.17 & 0.1 & $2430\pm600$ \\
804 & 686 Gersuind  &-0.12 &-0.64 & 5.3 & $800\pm80$    \\
805 & 945 Barcelona & 1.48 & 1.32 & 0.5 & $250\pm10$    \\
807 &4203 Brucato   & 0.29 & 0.01 &50.0 & $480\pm100$   \\
601 &  10 Hygiea    & 0.58 & 0.29 & 0.1 & $2420\pm580~(\dagger)$  \\
602 &  24 Themis    & 0.03 & 0.08 & 0.1 & $1500\pm320~(\dagger)$ \\
605 & 158 Koronis   & 0.99 & 1.06 & 0.1 & $2360\pm490~(\dagger)$ \\ 
606 & 221 Eos       &-0.60 &-0.60 & 0.1 & $1300\pm200~(\dagger)$ \\
607 & 283 Emma (s)  & 0.17 &-0.69 & 0.8 & $400\pm40$    \\
608 & 293 Brasilia  &-0.14 &-0.45 &17.7 & $160\pm10$    \\
609 & 490 Veritas   &-0.16 &-0.32 & 5.4 & $1820\pm180$  \\
610 & 832 Karin (s) &-0.29 & 0.89 & 1.0 & $5.8\pm0.2~(\dagger) $ \\
611 & 845 Naema     & 0.10 &-0.65 &19.2 & $210\pm10$    \\
612 &1400 Tirela    &-0.14 &-0.35 &21.9 & $1980\pm500$  \\
613 &3556 Lixiaohua & 1.30 &-0.38 &50.0 & $430\pm20$    \\
631 & 375 Ursula (s) & 0.06 & 0.70 & 4.3 & $1060\pm60~(\dagger)$ \\
901 &  31 Euphrosyne (s)&1.97&1.02 & 0.4 & $1380\pm70$   \\
\hline
\end{tabular}
\end{center}
\end{table*}
\twocolumn

\section{Conclusions}
\label{sec: conc}
The main results of this work can be summarized as follows:

\begin{itemize}

\item We developed a model for how the original field of velocities at 
infinity should appear for newly created families.  Families with a low  
$v_{EJ}/v_{esc}$ ratio, where $v_{EJ}$ is a parameter describing the standard 
deviation of ejection velocity field, assumed normal (Eq.~\ref{eq: std_veJ}), 
and $v_{esc}$ is the escape velocity from the parent body, should have 
values of velocities at infinity following a peaked distribution.
This distribution is characterized by positive values of kurtosis 
${\gamma}_2$ (leptokurtic distribution).  This is confirmed by 
SPH simulations for the Karin cluster, that show a leptokurtic 
distribution of $v_W$ values, the component of the velocity of infinity 
perpendicular to the orbital plane.  Since the proper inclination is less 
affected by dynamical evolution than proper semi-major axis and 
eccentricity, and since $v_W$ can be obtained from the distribution in 
proper $i$ (Eq.~\ref{eq: gauss_3}), we decided to concentrate our 
analysis on the distribution of $v_W$ values.  We studied the 
influence that dynamics has on the distribution of $v_W$ values.  An 
initially peaked $v_W$ distributions tend to become more 
normal with time.

\item We computed values of ${\gamma}_2(v_W(D_3))$, the kurtosis 
of the distribution of $v_W$ values for objects with $2 < D < 4$ ($D_3$), 
for all families listed in Nesvorn\'{y} et al. (2015) that i) have an 
age estimate, ii) have a $D_3$ population of at least 100 objects, 
and iii) are not following a Gaussian distribution, according to
the Jarque-Bera test.  We also required the families to have values of 
skewness, the parameter identifying possible asymmetries in the $v_W$ 
distribution, between -0.25 and 0.25.  Eight families, the Gallia, 
Barcelona, Hansa, Massalia, Koronis, Rafita, Hygiea and Dora families, 
satisfied these criteria and have values of ${\gamma}_2(v_W(D_3))$ larger than
0.25, which suggests that they could still bear traces of the original
values of $v_W$.  Four of these families are younger than 700 Myr.
All of these groups appear to be located in regions not strongly 
affected by dynamical mechanisms able to modify proper inclination.
Among asymmetrical families, the Karin and Astrid groups stand out as
interesting cases.

\end{itemize}

Overall, while a more in depth study of the families selected
in this work, accounting for a better analysis of local dynamics
and of the role of possible interlopers, should be performed, 
our analysis already allowed us to select several asteroid families
that could still bear traces of the original distribution
of velocities at infinity.  While to select families that could
still bear traces of the original values of $v_W$ we concentrated
on the shape of the $v_W$ distribution,  obtaining better estimates of the 
{\it magnitude} of $v_W$ for these selected families could much improve
our knowledge of the physical mechanisms at work in the 
family-forming event, and serve as an useful constraints for
simulations describing cratering and catastrophic destruction 
events, other than the observed Size Frequency Distribution 
(Durda et al. 2007).

\section*{Acknowledgments}

We thank an anonymous reviewer for comments and suggestions that
significantly improved the quality of this paper.  This paper was written 
while the first author was a visiting scientist at the Southwest Research 
Institute (SwRI) in Boulder, CO, USA.  We would like to thank the S\~{a}o 
Paulo State Science Foundation (FAPESP) that supported this work via the 
grant 14/24071-7, the Brazilian National Research Council (CNPq, grant 
305453/2011-4), and the National Science Foundation (NSF).  The authors are 
grateful to W. F. Bottke for allowing them to use the results of the Koronis 
family simulations, to S. Aljbaae for revising earlier 
version of this manuscript, and to D. Durda for comments and 
suggestions that improved the quality of this work.

\bsp

\label{lastpage}

\end{document}